\documentclass[twocolumn,showpacs,preprintnumbers,amsmath,amssymb,superscriptaddress,prl]{revtex4}
\usepackage{graphicx}% Include figure files
\usepackage{dcolumn}% Align table columns on decimal point
\usepackage{hyperref}
\usepackage{bm}% bold math

\begin{document}

%\preprint{APS/123-QED}

\title{Effects of intrinsic spin-relaxation in molecular magnets on
current-induced\\ magnetic switching}
 % Force line breaks with \\

\author{Maciej Misiorny}
 \email{misiorny@amu.edu.pl}
\affiliation{Department of Physics, Adam Mickiewicz University,
61-614 Pozna\'{n}, Poland}
%\textbf{}}%

\author{J\'{o}zef Barna\'{s}}
\email{barnas@amu.edu.pl} \affiliation{Department of Physics, Adam
Mickiewicz University, 61-614 Pozna\'{n}, Poland}
\affiliation{Institute of Molecular Physics, Polish Academy of
Sciences, 60-179 Pozna\'{n}, Poland
}%

\date{\today}% It is always \today, today,

\begin{abstract}
Current-induced magnetic switching of a single magnetic molecule
attached to two ferromagnetic contacts is considered theoretically,
with the main emphasis put on the role of intrinsic spin relaxation
processes. It is shown that spin-polarized current can switch
magnetic moment of the molecule, despite of the intrinsic spin
relaxation  in the molecule. The latter processes increase the
threshold voltage (current) above which the switching takes place.
\end{abstract}


%72.25.-b Spin polarized transport
%73.23.-b Electronic transport in mesoscopic systems
%75.50.Xx  Magnetic devices: molecular magnets
%75.60.Jk Magnetization reversal mechanisms
%85.75.-d Magnetoelectronics; spintronics: devices exploiting spin
%polarized transport or integrated magnetic fields

\maketitle

Single-molecule magnets
(SMMs)~\cite{Sessoli_Nature365/93,Gatteschi_book} attract much
attention due to their exceptional properties and possible
applications in quantum information
processing~\cite{Leuenberger_Science410/01} and information storage
technology~\cite{Sessoli_Nature365/93,Joachim_Nature408/00}. Apart
from this, SMMs are also promising as key elements of novel
spintronics devices~\cite{Rocha_NatureMat4/05}. Therefore, an
important question is how to manipulate the SMM in order to write a
bit of information on it. One possibility relies on the application
of an external magnetic field. In the following paper, however, we
consider another possibility, i.e. the current-induced magnetic
switching (CIMS)~\cite{Misiorny_PRB75/07,Misiorny_PRB76/07}. The
phenomenon of CIMS~\cite{Barnas_PRB72/05} is well known in the case
of artificial layered nanostructures. Since the present-day
technology allows to attach a SMM to electronic
contacts~\cite{Heersche_PRL96/06}, CIMS of a SMM is an alternative
way of writing information in SMM-based memory elements.

There are several challenging aspects of the current-induced
manipulation of SMM's spin. First, the up-to-date experimental
techniques offer only limited control of the relative orientation of
the molecule's easy axis and leads'
magnetizations~\cite{Timm_PRB76/07}. Second, intrinsic
spin-relaxation time of the molecule~\cite{Morello_PRL93/04} has a
significant influence on the switching parameters and is hardly
controllable externally.  Finally, the efficiency of spin injection
from ferromagnetic leads to molecules is a subject of intense
technological efforts. The main objective of this paper is a
detailed analysis of the second point, i.e. of the influence of
intrinsic spin-relaxation on the CIMS of a SMM.

\begin{figure}
    \includegraphics[width=0.9\columnwidth]{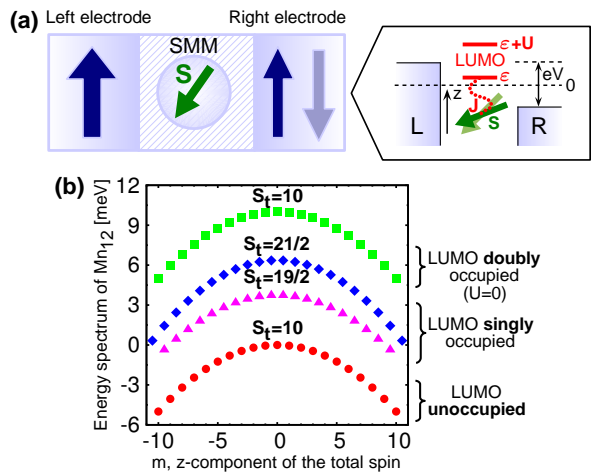}
    \caption{\label{Fig1} (color online)(a) Schematic representation of the system
    and switching mechanism due to spin-polarized current. (b) Energy
    levels of the $\textrm{Mn}_\textrm{12}$ molecule for the following
    parameters: $D\approx 0.05$ mV, $D_1\approx-0.006$ meV,
    $D_2\approx0.0017$ meV~\cite{Soler_JAmChemSoc125/03}, $J=0.25$ meV,
    $\varepsilon=5$ meV, and $U=0$. Different parabolas correspond to
    indicted values of the SMM's total spin $S_t$ and occupation numbers
    of the LUMO level.
    }
    \end{figure}

It is only very recently, when the switching of SMM's spin due to
spin-polarized current has been
proposed~\cite{Misiorny_PRB75/07,Misiorny_PRB76/07}. However, the
intrinsic spin relaxation in the molecule has not been taken into
account. When the energy $\varepsilon$ of the lowest unoccupied
orbital (LUMO) level of the molecule is sufficiently low,
electronic transport takes place owing to tunneling between the
electrodes and the LUMO level. The CIMS can then occur when the
LUMO level is exchange coupled to the SMM's spin. The
corresponding Hamiltonian of the molecule can be written in the
form
    \begin{multline}
    \mathcal{H}_{S\! M\! M} = -\big(D +\sum_{\sigma}D_{1}\,
    c_\sigma^\dag c_\sigma^{} + D_{2}\, c_\uparrow^\dag c_\uparrow^{}
    c_\downarrow^\dag c_\downarrow^{}
    \big)S_z^2
    \\
    + \sum_{\sigma}\varepsilon c_\sigma^\dag c_\sigma^{} + U
    c_\uparrow^\dag c_\uparrow^{} c_\downarrow^\dag c_\downarrow^{}
    -\frac{1}{2}
    \sum_{\sigma\sigma'}J\bm{\sigma}_{\sigma\sigma'}\cdot\bm{S}
    c_\sigma^\dag
    c_{\sigma'},
    \end{multline}
where ${\bm \sigma}$ is the Pauli spin  operator for electrons in
the LUMO level, $c_\sigma^{\dag} (c_\sigma^{})$ is the relevant
creation (anihilation) operator, and $U$ is the Coulomb energy of
two electrons of opposite spins in the LUMO level. The first term of
$\mathcal{H}_{S\! M\! M}$ describes the anisotropy of a SMM, whereas
the final one accounts for the exchange interaction between the
SMM's core and the LUMO level, with $J$ being the relevant exchange
parameter. The influence of molecule's oxidation state on the
anisotropy~\cite{Soler_JAmChemSoc125/03} is taken into account by
the terms linear in $D_{1}$ and $D_{2}$. In turn, tunneling
processes between the molecule and leads are described by
$\mathcal{H}_T$, $\mathcal{H}_T = \sum_{q}\sum_{{\bf
k}\sigma}\big[T_q a_{{\bf k}\sigma}^{q\dag}c_\sigma^{} + T_q^*
c_\sigma^\dag a_{{\bf k}\sigma}^q\big]$, where $T_q$ is the
tunneling matrix element between the SMM and the $q$-th lead ($q=L
(R)$ for the left (right) electrode), and $a_{{\bm k}\sigma}^{q}$
($a_{{\bm k}\sigma}^{q\dag}$) is the annihilation (creation)
operator of an electron with the wave vector ${\bm k}$ and spin
$\sigma$ in the $q$-th electrode. The system is shown schematically
in Fig.~\ref{Fig1}(a).

Tunneling between the leads and molecule gives rise to a finite
spin-dependent width $\Gamma_\sigma$ of the LUMO level,
$\Gamma_\sigma =\sum_q \Gamma_\sigma^q$, where $\Gamma_\sigma^q
=2\pi|T_q |^2D_\sigma^q$ and $D_\sigma^q$ is the spin-dependent
density of states (DOS) at the Fermi level in the lead $q$. The
parameters $\Gamma_\sigma^q$ will be used in the following to
describe coupling strength between the LUMO level and leads. It is
convenient to write $\Gamma_\sigma^q$ as
$\Gamma^{q}_{\pm}=\Gamma_{q}(1\pm P_q)$, where
$\Gamma_{q}=(\Gamma^{q}_{+}+\Gamma^{q}_{-})/2$, and $P_q$ is the
polarization of the $q$-th lead,
$P_q=(D_+^q-D_-^q)/(D_+^q+D_-^q)$. Here $\sigma = +(-)$
corresponds to spin-majority (spin-minority) electrons. In the
following, we assume that the couplings are symmetric,
$\Gamma_{\rm L}=\Gamma_{\rm R}= \Gamma/2$.

When the energy $\varepsilon$ of the LUMO level is large enough,
electron tunneling to the molecule is energetically forbidden at
bias voltages of interest. However, current still can flow due to
higher order processes, e.g. cotunneling ones, and CIMS of the
molecule's spin is still possible~\cite{Misiorny_PRB75/07} when
the electrons virtually entering the molecule couple to the
molecule's spin via the exchange interaction. The Hamiltonian of
the molecule can be then reduced to
$\mathcal{H}_{S\!M\!M}=-DS_z^2$, while tunneling processes can be
described effectively by
$\mathcal{H}_T=\frac{1}{2}\sum_{qq'}\sum_{\sigma\sigma'{\bm k}{\bm
k'}} (J {\bm \sigma}_{\sigma\sigma'}\cdot{\bm S}+\delta )a_{{\bm
k}\sigma}^{q\dag}a_{{\bm k'}\sigma'}^{q'}$, where $\delta$ takes
into account those tunneling processes between the leads, which
are not included in the exchange term. These, however, are
irrelevant from the point of view of switching process and can be
neglected ($\delta =0$).

Switching of the SMM's spin takes place consecutively {\it via}
the magnetic states of the molecule. These states are described by
the eigenvalue $m$ of the $z$ component of the molecule's total
spin, $S_t^z\equiv S_z+\frac{1}{2}\big(c_\uparrow^\dag
c_\uparrow^{}-c_\downarrow^\dag c_\downarrow^{}\big)$ (where the
second term represents the contribution from electrons in the LUMO
level), and the corresponding occupation number $n$ of the LUMO
level, i.e. $|n,m\rangle$ ~\cite{Misiorny_PRB76/07}.

The energy levels corresponding to the Hamiltonian $\mathcal{H}_{S\!
M\! M}$ form the four parabolas shown in Fig.~\ref{Fig1}(b). The
lowest parabola corresponds to the empty LUMO level and is the only
relevant energy spectrum in the limit of high LUMO level. In the
latter case, the spin reversal proceeds {\it via} all the
consecutive intermediate states $|0,-S+1\rangle...|0,S-1\rangle$
(dots in Fig.~\ref{Fig1}(b)). A more complicated situation occurs
when electrons tunnel directly through the LUMO level, and
transitions between neighboring molecular states are governed by the
following selection rules~\cite{Misiorny_PRB76/07}: $|\Delta
S_t^z|=1/2$ and the oxidation state of the SMM changes by one.
Consequently, direct transitions between molecular states of the
same spin multiplet are forbidden.

A convenient way of analyzing the magnetic switching of a SMM is to
consider the mean value of  the $z$ component of the total
molecule's spin,
   \begin{equation}
    \langle S_t^z\rangle = \sum_{n,m} mP_{|n,m\rangle}.
    \end{equation}
The problem is then reduced to determining the probabilities
$P_{|n,m\rangle}$ of finding the molecule in  all possible molecular
states $|n,m\rangle$. These probabilities can be determined from the
relevant master equations and the corresponding transition rates
between the molecular states. The key point is that these transition
rates must include also the effects due to intrinsic spin
relaxation.

Generally, in the systems under consideration one can distinguish
two classes of SMM's spin relaxation processes. The first class is
associated with the coupling of the molecule to ferromagnetic
leads~\cite{Misiorny_PRB75/07,Misiorny_PRB76/07,Misiorny_EPL78/07},
and the other one includes all intrinsic spin-relaxation
processes~\cite{Gatteschi_book,Morello_PRL93/04}. The role of the
latter processes in the CIMS of the SMM's spin is the main
objective of this paper. It is important to note that even at low
temperatures the molecule's spin is subject to decoherence due to
interaction with its environment. A SMM in an excited molecular
spin level can undergo transitions to neighboring levels of lower
energy, which is accompanied by emission of a phonon. As a
consequence, excited molecular spin states have a finite
life-time, and it has been shown that this time for
$\textrm{Fe}_\textrm{8}$ is of order of $10^{-6}$
s~\cite{Morello_PRL93/04}. Furthermore, coherence of the SMM's
spin can also be lost due to various forms of magnetic
interactions with the environment, e.g. due to the hyperfine
interaction with nuclear moments of protons in the vicinity of the
molecule~\cite{Gatteschi_book,Morello_PRL93/04}.

To include the intrinsic spin relaxation processes into
considerations, we introduce the relaxation rate $\gamma_R$ in
addition to the rates
$\gamma^{|n,m\rangle|n^\prime,m^\prime\rangle}$ describing
current-induced transitions between the molecular spin states
$|n,m\rangle$ and $|n^\prime,m^\prime\rangle$. The latter ones can
be calculated from the Fermi golden
rule~\cite{Misiorny_PRB75/07,Misiorny_PRB76/07}. In turn, intrinsic
relaxation of the molecule's spin occurs as transitions between
neighboring molecular states of the same spin multiplet,
Fig.~\ref{Fig1}(b), i.e. the occupation of the LUMO level is not
changed by these processes. Furthermore, we assume that the spin
relaxation is fully characterized by a phenomenological relaxation
time $\tau_R$, i.e. the relaxation rate takes the form
\begin{equation}\label{eq:relax_rates}
    \gamma_R^{|n,m\rangle|n,m^\prime\rangle}=\frac{1}
    {\tau_R}\times\frac{\exp\Big[\frac{\Delta}{2k_B T}\Big]}
    {2\cosh\Big[\frac{\Delta}{2k_B T}\Big]}.
    \end{equation}
Here, $\epsilon_{|n,m\rangle}$ denotes the energy of the molecular
state $|n,m\rangle$, $k_{\rm B}$ is the Boltzmann constant, $T$ is
the temperature of the system, and
$\Delta=\epsilon_{|n,m\rangle}-\epsilon_{|n,m^\prime\rangle}$. The
Boltzmann factor in Eq.~(\ref{eq:relax_rates}) assures that the
intrinsic spin relaxation drives the SMM's spin to the state of the
lowest energy.

Taking into account the relaxation processes discussed above, the
master equations for the probabilities $P_{|n,m\rangle}$ take the
form,
   \begin{multline}
    c\frac{dP_{|n,m\rangle}}{dV}= -\Big(\gamma_R^{|n,m\rangle|n,m-1\rangle} +
    \gamma_R^{|n,m\rangle|n,m+1\rangle}
    \Big)P_{|n,m\rangle}
    \\
    +\gamma_R^{|n,m-1\rangle|n,m\rangle}P_{|n,m-1\rangle}
    +\gamma_R^{|n,m+1\rangle|n,m\rangle}P_{|n,m+1\rangle}
    \\
    +\sum_{n',m^\prime}\Big[
    \gamma^{|n',m^\prime\rangle|n,m\rangle}P_{|n',m^\prime\rangle}-
    \gamma^{|n,m\rangle|n',m^\prime\rangle}
    P_{|n,m\rangle}\Big].
    \end{multline}
In the following we assume that initially the molecule is saturated
in the state $|0,-10\rangle$, and then voltage growing linearly in
time is applied,  $V=ct$, with  $c$ denoting the speed at which the
voltage is augmented. It means that for the molecule of the spin
$S=10$, like the molecule $\textrm{Mn}_\textrm{12}$ or
$\textrm{Fe}_\textrm{8}$, one has to solve the set of 21 coupled
differential equations for the situation of large LUMO level and 84
equations in the general case.

    \begin{figure}
    \includegraphics[width=0.7\columnwidth]{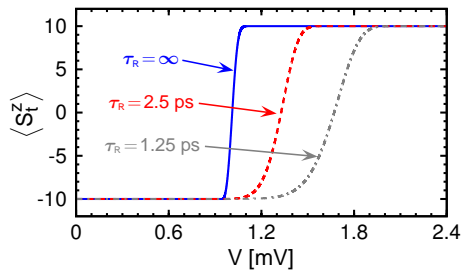}
    \caption{\label{Fig2} (color online) The effect of intrinsic relaxation
    processes on magnetic switching of the molecule
    $\textrm{Mn}_\textrm{12}$ in the limit of high LUMO level,
    calculated for indicated values of the relaxation time,
    $\tau_R$, and for parallel magnetic configuration. The
    polarization parameters of the electrodes are: $P_L=1$ and
    $P_R=0.5$. The other parameters are: $J=100$ meV, $D\approx 0.05$
    meV~\cite{Soler_JAmChemSoc125/03}, $T=0.01$ K, and $c=10$ kV/s.
    }
    \end{figure}

In Fig.~\ref{Fig2} we show evolution of the $z$ component of the
molecule's spin in the case of parallel magnetic configuration and
high LUMO level (current flows then due to higher order
processes). The results clearly show that the molecule's spin
becomes switched when the voltage exceeds some critical value,
which is determined by the magnetic anisotropy (energy gap between
the states corresponding to $m=-10$ and $m=-9$) and the intrinsic
relaxation time. Since the intrinsic spin-flip relaxation
processes tend to restore the initial state, the lowest threshold
voltage occurs in the absence of intrinsic spin relaxation. The
switching, however, takes also place in the presence of intrinsic
spin relaxation processes, although the threshold voltage becomes
increased. Apart from this, the switching time also increases with
decreasing $\tau_R$. Similar behavior also occurs in the case when
magnetic moments of the leads are antiparallel.

    \begin{figure}
    \includegraphics[width=0.6\columnwidth]{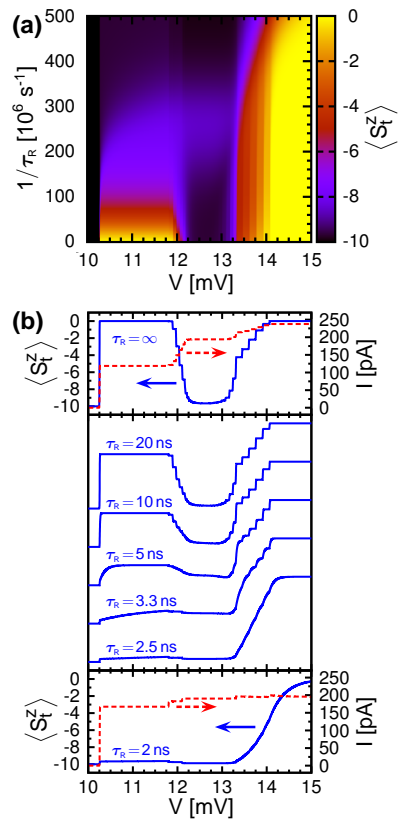}
    \caption{\label{Fig3} (color online) (a) The mean value of the total spin
    $\langle S_t^z\rangle$ for different values of the inverse
    relaxation time, $\tau_R^{-1}$,  in the case of parallel
    configuration of the electrodes' magnetic moments, calculated for
    $P_L=P_R=0.5$. Solid lines in the part (b) represent cross sections
    of the plot (a) for several values of $\tau_R$, whereas the dashed
    lines correspond to the current flowing through the system. The
    other parameters are as in Fig.~1, and $T=0.01$ K, $c=1$ V/s, and
    the coupling parameter $\Gamma=0.001$ meV.
    }
    \end{figure}

    \begin{figure}
    \includegraphics[width=0.6\columnwidth]{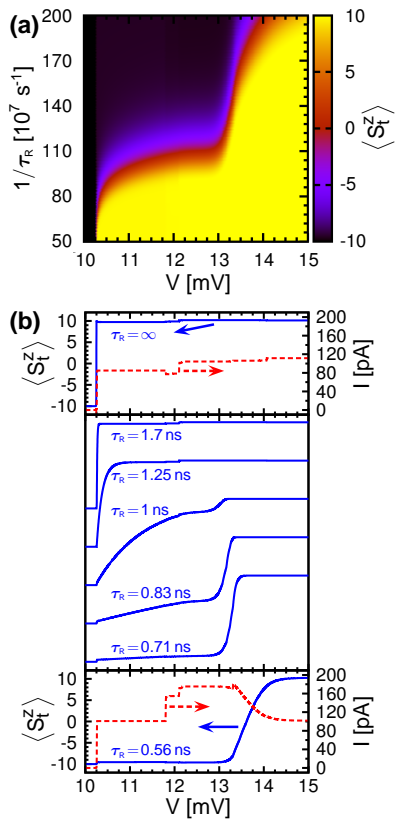}
    \caption{\label{Fig4} (color online) The same as in Fig.~3 but for antiparallel
    alignment of the electrodes' magnetic moments.
    }
    \end{figure}

The parameters assumed in Fig.~\ref{Fig2} correspond to
half-metallic ferromagnetic left electrode ($P_L=1$), and typical
3d ferromagnetic metallic right electrode. For simplicity the
positive bias corresponds to electrons flowing from left to right
($e>0$), i.e. from half-metallic ferromagnetic electrode to the 3d
one. Spin-up electrons leaving the half-metallic electrode can
change its spin orientation when interacting {\it via} exchange
coupling with the molecule's spin, and this way can increase the
spin number $m$ of the molecule's spin. Intrinsic relaxation
processes tend to restore the initial state. When the current
exceeds some critical value, the competition of intrinsic spin
relaxation (lowering the quantum number $m$) and current-induced
processes (increasing the number $m$) leads to spin reversal of
the molecule. This takes place in both, parallel and antiparallel
(with magnetic moment of the right electrode being reversed)
magnetic configurations. For reversed bias polarization only
switching from the state $|0,10\rangle$ to the state
$|0,-10\rangle$ is possible.

In Figs~\ref{Fig3} and \ref{Fig4} we show the average value of the
total spin $\langle S_t^z\rangle$ and current flowing in a biased
system in the case when switching occurs due to sequential
tunneling of electrons through the molecule's LUMO level. These
two figures correspond to parallel (Fig.~\ref{Fig3}) and
antiparallel (Fig.~\ref{Fig4}) magnetic configurations. Clearly,
there is no switching in the parallel configuration. Instead of
this, current excites the molecule to higher states and the
average spin becomes zero (see Fig.~\ref{Fig3}). The situation is
different in the antiparallel configuration, where there is a
clear switching from the state $|0,-10\rangle$ to the state
$|0,10\rangle$. To understand this behavior one should note that
in Figs~\ref{Fig3} and \ref{Fig4} the spin polarization of both
electrodes is the same. Consequently, the current-induced
processes increasing the number $m$ and those decreasing $m$ occur
with the same rate in the parallel configuration. Accordingly,
none of the molecule's spin states is stabilized by the current.
In contrast, in the antiparallel configuration processes
increasing the number $m$ start to dominate over those decreasing
$m$ above a certain threshold voltage, and the switching to the
state $|0,10\rangle$ takes place. Current-induced switching of the
molecules's spin may be possible also in the parallel
configuration, provided spin polarizations of the electrodes are
different.

In conclusion, we have shown that spin-polarized current flowing
though the molecule can switch its magnetic moment despite of
intrinsic spin relaxation processes in the molecule. The latter
processes increase the threshold voltage (current) and switching
time. If for a certain bias polarization current stabilizes the
state $|0,-10\rangle$ (or $|0,10\rangle$), then the opposite current
stabilizes the state $|0,10\rangle$ (or $|0,-10\rangle$).

\emph{Acknowledgments} This work, as part of the European Science
Foundation EUROCORES Programme SPINTRA, was supported by funds from
the Ministry of Science and Higher Education as a research project
in years 2006-2009 and the EC Sixth Framework Programme, under
Contract N. ERAS-CT-2003-980409.

\end{document}